\documentclass[onecolumn,showpacs,preprintnumbers,amsmath,amssymb]{revtex4}

\usepackage{graphicx}% Include figure files
\usepackage{dcolumn}% Align table columns on decimal point
\usepackage{bm}% bold math
\begin{document}
\title{Photon wave functions and quantum interference experiments}
\author{G.G. Lapaire and J.E. Sipe}
\address{Department of Physics and Institute for Optical Sciences,\\
University of Toronto, 60 St. George Street,\\
Toronto\\
ON M5S 1A7 CANADA}

\begin{abstract}
We present a general theory to describe two-photon interference, including a
formal description of few photon intereference in terms of single-photon
amplitudes. With this formalism, it is possible to describe both frequency
entangled and separable two-photon interference in terms of single-photon
wave functions. Using this description, we address issues related to the
physical interpretation of two-photon interference experiments. We include a
discussion on how few-photon interference can be interpreted as a bosonic
exchange effect, and how this relates to traditional exchange effects with
fermions.
\end{abstract}
\maketitle

Newton and Wigner\cite{NW} first discussed the non-localizability of
photons, which prevents the introduction of a position-representation wave
function in the usual sense of a wave function for a massive particle.
Nonetheless, it has been known for years that it is possible to present
physically meaningful descriptions of photon detection in finite regions of
space\cite{MANDELETC}. This has led to a host of approaches to introduce a
\textquotedblleft photon wave function\textquotedblright\ \cite{SIPEBIAL},
each of which establishes at least a limited analogy to massive particle
wave functions. Today there is no single accepted definition of such a
photon wave function, as there are a variety of possible analogies, and the
most convenient often depends on the system one would like to describe. For
example, Chan {\it et al.} \cite{CHAN} discuss a wave function associated
with a photon spontaneously emitted from an atom in terms Schmidt pairs of
atomic and photonic eigenfunctions, whereas Resch {\it et al.} \cite{RESCH} find it useful to consider a
photon wave function based on the Glauber detection probability\cite{MW} in understanding an absorptive
exchange effect.
Given the variety of convenient types of \textquotedblleft photon wave
functions\textquotedblright\ introduced in the literature, it is perhaps
best not to insist on a particular definition, but rather to understand the
term to refer broadly to any approach for describing a photon in a manner
analogous to the usual massive particle wavefunctions introduced in
nonrelativistic physics. That is the point of view we take here, where we
use photon wave functions to describe few-photon interference experiments.
Although in the particulars of our discussion we use a photon wave function
definition based on the Glauber detection formula, within the usually
relevant approximations an easy translation into other photon wave function
definitions could be made.

Yet one might ask, \textquotedblleft why bother?\textquotedblright\ After
all, the measurement results of two-photon interference experiments can be
predicted with relatively simple, well-known calculations. Still, what seems
to be a less straightforward task is the physical interpretation of the
experiments. Few-photon interference is often discussed loosely in terms of
interfering Feynman paths, or the overlap of wave packets, or the
distinguishability of a particular set of outcomes. In this paper we
formally address what one must mean if one wishes to discuss the
interference of individual photon amplitudes leading to a particular
detection event. Using photon wave functions, we explicitly show that
second-order interference experiments can be understood in terms of
single-photon amplitudes. Consequently, this approach yields considerable
insight into few-photon interference experiments, and illustrates the
relevance of the bosonic nature of the photon\cite{FEARN1} by contrasting
the corresponding description that would result for fermions. It is with a
photon wave function description that one can perhaps best isolate
non-classical interference terms, and discuss the interference of
single-photon amplitudes corresponding to a particular detection.

Perhaps the most familiar few-photon interference experiment involves the
Hong-Ou-Mandel interferometer\cite{HOM}. We show how second-order
interference of this type manifests itself as an exchange effect in the
photon wave function picture. This seems to be appreciated by many workers
in the field, although we have not been able to find an explicit discussion
of this point in the literature. We then move on to other experiments. We
show how the elimination of which-path distinguishing information restores
the exchange effect in quantum eraser experiments\cite{QE}. And while in
their discussion of a postponed compensation experiment\cite{PITTMAN1}
Pittman {\it et al.} emphasize the limitations of using a pair of photon
wave packets to describe frequency entangled two-photon interference, we
show that it is indeed possible to understand this experiment in terms of
the interference of single photon amplitudes. Before proceeding to these
issues in section II below, we identify our definition and notation for
few-photon wave functions in section I; our conclusions are presented in
section III.

\section{Photon wave functions}

\label{wavefunc}Photon wave functions can be extracted from the usual field
theory description of photodetection, and used in a way similar to the use
of {\it massive} particle wave functions in describing the detection of
those particles. To establish this analogy we recall some well-known results
from the field theory for nonrelativistic massive particles, in which an
arbitrary single-particle state $\left\vert S\right\rangle $ can be written
as 
\begin{equation}
\left\vert S\right\rangle =\sum_{\lambda }g^{S}(\lambda )a^{\dagger }\left(
\lambda \right) \left\vert vac\right\rangle ,  \label{eq:massive-single}
\end{equation}%
where each $\lambda $ identifies one of a set of normalized, orthogonalized
single-particle modes, $\left\vert vac\right\rangle $ is the vacuum state,
and $a^{\dagger }\left( \lambda \right) $ is the associated particle
creation operator. For particles in free space we can take $\lambda =(\mu ,%
{\bf k})$ to label both the spin state $\mu $ and the wave vector ${\bf k}$
of a plane wave. Here $g^{S}(\lambda )$ is a normalized amplitude,
satisfying 
\begin{equation}
\sum_{\lambda }\left\vert g^{S}(\lambda )\right\vert ^{2}=1.  \label{gsnorm}
\end{equation}%
We can write this state (\ref{eq:massive-single}) in terms of the creation
operator for a particle with spin label $\mu $ at ${\bf r}$, 
\[
\widehat{\Psi }_{\mu }^{\dagger }\left( {\bf r}\right) =\frac{1}{\sqrt{V}}%
\sum_{{\bf k}}a^{\dagger }\left( \lambda \right) e^{i{\bf k\cdot r}} 
\]%
where $V$ is a normalization volume, as 
\begin{equation}
\left\vert S\right\rangle =\sum_{\mu }\int d{\bf r\,}G_{\mu }^{S}({\bf r})%
\widehat{\Psi }_{\mu }^{\dagger }\left( {\bf r}\right) \left\vert
vac\right\rangle ,  \label{eq:massive-single2}
\end{equation}%
where 
\[
G_{\mu }^{S}({\bf r})=\frac{1}{\sqrt{V}}\sum_{{\bf k}}g^{S}(\lambda )e^{i%
{\bf k\cdot r}}, 
\]%
and we anticipate the passage to the limit of infinite $V$ and the resulting
continuous range of ${\bf k}$ over plane wave modes. We can now identify the
single-particle wave function $\Phi _{\mu }^{S}({\bf r},t)$ associated with
the single particle state $\left\vert S\right\rangle $. The state $\widehat{%
\Psi }_{\mu }\left( {\bf r,}t\right) \left\vert S\right\rangle $, where $%
\widehat{\Psi }_{\mu }\left( {\bf r,}t\right) =$ $\widehat{U}\left(
t,0\right) \widehat{\Psi }_{\mu }\left( {\bf r}\right) \widehat{U}^{\dagger
}\left( t,0\right) $ and $\widehat{U}\left( t,0\right) $ is the time
evolution operator from $0$ to $t$ determined by the Schr\"{o}dinger
equation, is proportional to the vacuum state; the wave usual function $\Phi
_{\mu }^{S}({\bf r},t)$ of elementary quantum mechanics provides just that
proportionality, 
\begin{equation}
\widehat{\Psi }_{\mu }\left( {\bf r,}t\right) \left\vert S\right\rangle
=\Phi _{\mu }^{S}\left( {\bf r,}t\right) \left\vert vac\right\rangle ,
\label{eq:massive-single3}
\end{equation}%
as can be easily confirmed. The probability of detecting the particle within 
$d{\bf r}$ of ${\bf r}$ at time $t$ can now be written in either the
particle or field theory notation,

\begin{equation}
\sum_{\mu }\left| \Phi _{\mu }^{S}\left( {\bf r,}t\right) \right| ^{2}d{\bf r%
}=\sum_{\mu }\left\langle S\right| \widehat{\Psi }_{\mu }^{\dagger }\left( 
{\bf r,}t\right) \widehat{\Psi }_{\mu }\left( {\bf r,}t\right) \left|
S\right\rangle \,d{\bf r}.  \label{eq:massive4}
\end{equation}
Moving to two-particle states, we can construct the most general such state $%
\left| T\right\rangle $ according to

\begin{equation}
\left\vert T\right\rangle =N^{T}\sum_{\lambda ,\lambda ^{\prime
}}f^{T}\left( \lambda ,\lambda ^{\prime }\right) a^{\dagger }\left( \lambda
\right) a^{\dagger }\left( \lambda ^{\prime }\right) \left\vert
vac\right\rangle =\sum_{\mu ,\mu ^{\prime }}\int \int d{\bf r}d{\bf r}%
^{\prime }F_{\mu \mu ^{\prime }}^{T}\left( {\bf r},{\bf r}^{\prime }\right) 
\widehat{\Psi }_{\mu }^{\dagger }\left( {\bf r}\right) \widehat{\Psi }_{\mu
^{\prime }}^{\dagger }\left( {\bf r}^{\prime }\right) \left\vert
vac\right\rangle ,  \label{eq:massive5}
\end{equation}%
where for later convenience we have introduced an explicit normalization
constant $N^{T}$ such that $\left\langle T|T\right\rangle =1$, and 
\[
F_{\mu \mu ^{\prime }}^{PQ}\left( {\bf r},{\bf r}^{\prime }\right) =\frac{%
N^{T}}{V}\sum_{{\bf k,k}^{\prime }}f^{T}\left( \lambda ,\lambda ^{\prime
}\right) e^{i{\bf k\cdot r}}e^{i{\bf k^{\prime }\cdot r}^{\prime }}. 
\]%
Without loss of generality the amplitude $f^{T}\left( \lambda ,\lambda
^{\prime }\right) $ can be taken to be symmetric with respect to the
interchange of $\lambda $ and $\lambda ^{\prime }$ for bosons, and
antisymmetric with respect to that interchange for fermions. A two-time,
two-particle wave function (or bi-particle wave function), $\Phi _{\mu \mu
^{\prime }}^{T}\left( {\bf r},{\bf r}^{\prime },t,t^{\prime }\right) $, can
be introduced according to

\begin{equation}
\frac{1}{\sqrt{2}}\widehat{\Psi }_{\mu }\left( {\bf r,}t\right) \widehat{%
\Psi }_{\mu ^{\prime }}\left( {\bf r}^{\prime }{\bf ,}t^{\prime }\right)
\left\vert T\right\rangle =\Phi _{\mu \mu ^{\prime }}^{T}\left( {\bf r},{\bf %
r}^{\prime },t,t^{\prime }\right) \left\vert vac\right\rangle .
\label{eq:massive6}
\end{equation}%
It is easy to show that $\Phi _{\mu \mu ^{\prime }}^{T}({\bf r,r}^{\prime
},t,t)$ is the usual two-particle wave function at time $t$, symmetric (or
antisymmetric) with the interchange of $({\bf r,}\mu )$ and $({\bf r}%
^{\prime },\mu ^{\prime })$, if we deal with bosons (or fermions). And
considering detection processess activated at time $t$, a standard
calculation easily done at the wave function level shows that the
probability of detecting one particle within $d{\bf r}$ of ${\bf r}$ and a
second particle within $d{\bf r}^{\prime }$ of ${\bf r}^{\prime }$ is given
by \cite{GRAHAM} 
\begin{equation}
\sum_{\mu \mu ^{\prime }}\left\vert \Phi _{\mu \mu ^{\prime }}^{T}\left( 
{\bf r},{\bf r}^{\prime },t,t\right) \right\vert ^{2}d{\bf r}d{\bf r}%
^{\prime }=\frac{1}{2}\sum_{\mu \mu ^{\prime }}\left\langle T\left\vert 
\widehat{\Psi }_{\mu ^{\prime }}^{\dagger }\left( {\bf r}^{\prime }{\bf ,}%
t\right) \widehat{\Psi }_{\mu }^{\dagger }\left( {\bf r,}t\right) \widehat{%
\Psi }_{\mu }\left( {\bf r,}t\right) \widehat{\Psi }_{\mu ^{\prime }}\left( 
{\bf r}^{\prime }{\bf ,}t\right) \right\vert T\right\rangle d{\bf r}d{\bf r}%
^{\prime }.  \label{eq:massive7}
\end{equation}

We now introduce photon wave functions in such a way that the equivalences (%
\ref{eq:massive4},\ref{eq:massive7}) between wave functions and field theory
descriptions hold when the standard Glauber detection formulas are used to
model photodetection probabilities. Between detector activations we assume
that the electromagnetic field evolves as a free radiation field. A modified
version of this approach can be written down if this is not the case, but we
will not do so here. We begin by introducing a single-photon state by
analogy with (\ref{eq:massive-single}), 
\begin{equation}
\left\vert S\right\rangle =\sum_{\lambda }g^{S}(\lambda )a^{\dagger }\left(
\lambda \right) \left\vert vac\right\rangle  \label{eq:photonp}
\end{equation}%
where $a^{\dagger }\left( \lambda \right) $ is the photon creation operator
for a mode $\lambda $, where the index $\lambda $ labels both the
polarization (or helicity) and the wave vector ${\bf k}$, and the amplitudes 
$\left\{ g^{S}(\lambda )\right\} $ again satisfy (\ref{gsnorm}). The
single-photon wave function at position ${\bf r}$ and time $t$, ${\bf \Phi }%
^{S}({\bf r,}t{\bf )}$ is defined by 
\begin{equation}
{\bf E}^{+}({\bf r},t)\left\vert S\right\rangle ={\bf \Phi }^{S}({\bf r}%
,t)\left\vert vac\right\rangle  \label{eq:singlewavefunction}
\end{equation}%
where 
\begin{eqnarray}
{\bf E}^{+}({\bf r},t) &=&i\sum_{\lambda }K_{k}\hat{e}_{\lambda }a\left(
\lambda \right) e^{i({\bf k\cdot r}-\omega _{k}t)},
\label{electricfieldcomponents} \\
{\bf E}^{-}({\bf r},t) &=&\left( {\bf E}^{+}({\bf r},t)\right) ^{\dagger } 
\nonumber
\end{eqnarray}%
are respectively the positive and negative frequency components of the
electric field operator ${\bf E(r},t)=$ ${\bf E}^{+}({\bf r},t)+{\bf E}^{-}(%
{\bf r},t),$ $K_{k}=\sqrt{2\pi \hbar \omega _{k}/V}$ with $V$ as the
normalization volume, $\hat{e}_{\lambda }$ is the unit vector indicating the
polarization, and $\omega _{k}=c\left\vert {\bf k}\right\vert $ is the
angular frequency of wave vector ${\bf k}$. The analogy of this photon wave
function with the single particle wave function of a nonrelativistic
particle lies in the fact that, for the single photon state, the Glauber
detection rate of a single ideal detector at position ${\bf r}$ and at time $%
t$ is proportional to the first order correlation function, given by\cite{MW}
\begin{equation}
w^{(1)}({\bf r},t)=\left\langle S\right\vert {\bf E}^{-}({\bf r},t)\cdot 
{\bf E}^{+}({\bf r},t)\left\vert S\right\rangle =\left\vert {\bf \Phi }^{S}(%
{\bf r},t)\right\vert ^{2},  \label{firstcorr}
\end{equation}%
{\it cf. }(\ref{eq:massive4}). For photon wave functions this is, of course,
not a fundamental postulate of the theory, but is derived from the Glauber
detection model; in particular, we use a form of the model where the
detector is assumed isotropic. The dynamics of ${\bf \Phi }^{S}({\bf r},t)$
are not governed by the Schr\"{o}dinger equation, as in the massive particle
case, but rather by Maxwell's equations.

The analogy with massive particles carries on to states with more than one
excitation. A general two-photon state of the radiation field is described
by 
\begin{equation}
\left\vert T\right\rangle =N^{T}\sum_{\lambda ,\lambda ^{\prime
}}f^{T}\left( \lambda ,\lambda ^{\prime }\right) a^{\dagger }\left( \lambda
\right) a^{\dagger }\left( \lambda ^{\prime }\right) \left\vert
vac\right\rangle  \label{eq:twophot}
\end{equation}%
where $N^{T}$ is the normalization factor, which for photons (or massive
bosons) can be taken to be 
\begin{equation}
N^{T}=\left( \sum_{\lambda ,\lambda ^{\prime }}\left\vert f^{T}\left(
\lambda ,\lambda ^{\prime }\right) \right\vert ^{2}+\sum_{\lambda ,\lambda
^{\prime }}\left[ f^{T}\left( \lambda ^{\prime },\lambda \right) \right]
^{\ast }f^{T}\left( \lambda ,\lambda ^{\prime }\right) \right) ^{-\frac{1}{2}%
},  \label{normalizationfactor}
\end{equation}%
to guarantee $\left\langle T|T\right\rangle =1.$ We do not choose the
function $f^{T}$ itself to be normalized, as it is convenient to write the
normalization factor separately when comparing two-photon states to
corresponding single-photon states. A special case is that for which the
amplitude $f^{T}\left( \lambda ,\lambda ^{\prime }\right) $ in some basis
(taken here to be that of polarizations or helicities and wave vectors) can
be written as the symmetric product of a function of $\lambda $ and a
function of $\lambda ^{\prime }$. Each of these functions can then be
associated with a one-photon state. In such a case we write $f^{T}(\lambda
,\lambda ^{\prime })$ as the function $f^{PQ}\left( \lambda ,\lambda
^{\prime }\right) =g^{P}(\lambda )g^{Q}(\lambda ^{\prime })+g^{P}(\lambda
^{\prime })g^{Q}(\lambda ),$ where $\left\{ g^{P}(\lambda )\right\} $ and $%
\left\{ g^{Q}(\lambda )\right\} $ are normalized spectral amplitudes (\ref%
{gsnorm}) associated with single photon states, and we call the two-photon
state {\it separable. }As is usually done in the literature, one can
describe the same state with the simpler non-symmetric function $%
f^{PQ}\left( \lambda ,\lambda ^{\prime }\right) =g^{P}(\lambda
)g^{Q}(\lambda ^{\prime })$ since the additional component does not change
the state (\ref{eq:twophot}). \ With this choice of $f^{PQ}\left( \lambda
,\lambda ^{\prime }\right) $, for such separable states,

\begin{equation}
N^{PQ}=\left( 1+\left| \sum_{\lambda }\left[ g^{P}(\lambda )\right]
^{*}g^{Q}(\lambda )\right| ^{2}\right) ^{-\frac{1}{2}},  \label{sepnorm}
\end{equation}
and $N^{PQ}=1$ only if the two single photon amplitudes are orthogonal; if $%
P=Q$ we have a state of two identical photons, and $N^{PQ}=2^{-1/2}$.

We introduce a two-photon wave function (sometimes called a biphoton\cite%
{RUBIN1}) that satisfies 
\begin{equation}
\frac{1}{\sqrt{2}}E_{j}^{+}({\bf r}_{2},t_{2})E_{i}^{+}({\bf r}%
_{1},t_{1})\left\vert T\right\rangle =\Phi _{_{ij}}^{T}({\bf r}_{1},{\bf r}%
_{2},t_{1},t_{2})\left\vert vac\right\rangle ,  \label{eq:twophotsat}
\end{equation}%
where Roman subscripts denote Cartesian components, and $\Phi _{_{ij}}^{T}(%
{\bf r}_{1},{\bf r}_{2},t_{1},t_{2})$ are the components of the wave
function associated with the state $\left\vert T\right\rangle $. In a
general two-photon state, the coincidence detection rate of two ideal
detectors at positions ${\bf r}_{1}$, ${\bf r}_{2}$ and at times $t_{1}$, $%
t_{2}$ is proportional to the second order correlation function\cite{MW} 
\begin{eqnarray}
w^{\left( 2\right) }({\bf r}_{1},{\bf r}_{2},t_{1},t_{2})
&=&\sum_{ij}\left\langle T\left\vert E_{i}^{-}({\bf r}_{1},t_{1})E_{j}^{-}(%
{\bf r}_{2},t_{2})E_{j}^{+}({\bf r}_{2},t_{2})E_{i}^{+}({\bf r}%
_{1},t_{1})\right\vert T\right\rangle  \label{secondcorr} \\
&=&2\sum_{ij}\left\vert \Phi _{_{ij}}^{T}({\bf r}_{1},{\bf r}%
_{2},t_{1},t_{2})\right\vert ^{2},  \nonumber
\end{eqnarray}%
{\it cf.} (\ref{eq:massive7}), where $\Phi _{ij}^{T}({\bf r}_{1},{\bf r}%
_{2},t_{1},t_{2})$ is symmetric under exchange of $(i,{\bf r}_{1},t_{1})$
with $(j,{\bf r}_{2},t_{2})$.

\section{Interference experiments}

\smallskip We now examine a type of quantum interference, observed in a
number of few-photon experiments, that is associated with the measurement of
particular coincidence detection rates. We discuss these rates in terms of
interfering single-photon amplitudes, and show that the interference can be
associated with photon exchange effects.

The simplest type of two-photon state is a separable one, in which we can
write $f^{PQ}\left( \lambda ,\lambda ^{\prime }\right) =g^{P}(\lambda
)g^{Q}(\lambda ^{\prime })$. Here we can say that one photon has the
spectral properties of `Photon $P$' and one photon has the properties of
`Photon $Q$' referring to the single-photon states $\left\vert
P\right\rangle $ and $\left\vert Q\right\rangle $. For massive fermions, the
analogous kind of separable spectral amplitude corresponds to a standard
Hartree-Fock, single-determinant wave function. In many-body physics one
usually characterizes such a state as free of the \textquotedblleft
correlation effects\textquotedblright\ that arise due to electron-electron
interactions in more sophisticated models of the full, many-electron wave
function. Nonetheless, there are dynamical consequences due to exchange
effects even in separable states, which arise for bosons as well, and in
particular photons. After considering the separable case, we generalize the
discussion to address frequency entangled few-photon interference
experiments that explicitly involve correlation effects.

The two-photon wave function for a separable two-photon state takes the form 
\begin{equation}
\Phi _{_{ij}}({\bf r}_{1},{\bf r}_{2},t_{1},t_{2})=\frac{N^{PQ}}{\sqrt{2}}%
\left[ \Phi _{_{i}}^{P}({\bf r}_{1},t_{1})\Phi _{_{j}}^{Q}({\bf r}%
_{2},t_{2})+\Phi _{_{i}}^{Q}({\bf r}_{1},t_{1})\Phi _{_{j}}^{P}({\bf r}%
_{2},t_{2})\right] ,  \label{eq:twophotwavesep}
\end{equation}%
with the $+$ arising instead of a $-$ because photons are bosons, and where $%
\Phi _{i}^{P}({\bf r},t)$ and $\Phi _{j}^{Q}({\bf r},t)$ are the
single-particle wave functions associated with photons $P$ and $Q$
respectively. It is in this form that one can discuss interference in terms
of individual photon amplitudes. The second-order correlation function is 
\begin{eqnarray}
w^{\left( 2\right) }({\bf r}_{1},{\bf r}_{2},t_{1},t_{2})
&=&\sum_{ij}\left\vert N^{PQ}\right\vert ^{2}\bigg[\left\vert \Phi
_{_{i}}^{P}({\bf r}_{1},t_{1})\right\vert ^{2}\left\vert \Phi _{_{j}}^{Q}(%
{\bf r}_{2},t_{2})\right\vert ^{2}+\left\vert \Phi _{_{j}}^{Q}({\bf r}%
_{1},t_{1})\right\vert ^{2}\left\vert \Phi _{_{i}}^{P}({\bf r}%
_{2},t_{2})\right\vert ^{2}  \nonumber \\
&&{}+\left[ \Phi _{_{i}}^{P}({\bf r}_{1},t_{1})\right] ^{\ast }\Phi
_{_{i}}^{Q}({\bf r}_{1},t_{1})\left[ \Phi _{_{j}}^{Q}({\bf r}_{2},t_{2})%
\right] ^{\ast }\Phi _{_{j}}^{P}({\bf r}_{2},t_{2})  \label{eq:exchangefec}
\\
&&{}+\left[ \Phi _{_{i}}^{Q}({\bf r}_{1},t_{1})\right] ^{\ast }\Phi
_{_{i}}^{P}({\bf r}_{1},t_{1})\left[ \Phi _{_{j}}^{Q}({\bf r}_{2},t_{2})%
\right] ^{\ast }\Phi _{_{j}}^{Q}({\bf r}_{2},t_{2}){}\bigg].  \nonumber
\end{eqnarray}%
The first two terms on the right-hand side of (\ref{eq:exchangefec}) are the
classical independent-particle terms. If only these terms were present, the
detection coincidence would be simply identified with the alternatives
\textquotedblleft photon $P$ at $({\bf r}_{1},t_{1})$ and photon $Q$ at $(%
{\bf r}_{2},t_{2})$\textquotedblright\ or \textquotedblleft photon $Q$ at $(%
{\bf r}_{1},t_{1})$ and photon $P$ at $({\bf r}_{2},t_{2})$%
\textquotedblright\ characteristic of independent detection events. The last
two terms on the right-hand side are the interference or exchange terms. One
can see that the exchange terms are proportional to the
indistinguishability, or overlap, of the $P$ and $Q$ single-photon
amplitudes in the region of interest. Non-zero exchange terms are indicative
of the \textquotedblleft indistinguishable Feynman paths\textquotedblright\
or \textquotedblleft overlapping single-photon amplitudes\textquotedblright\
sometimes mentioned in the few-photon literature.

It is important to consider the finite detection window of a realistic
detector. Though the exchange terms in (\ref{eq:exchangefec}) may be
non-zero for some values of $t_{1}$ and $t_{2}$, the exchange terms may
integrate to zero over a finite detection time. Particularity if one were to
consider photons in different frequency ranges, one would find that
spatially overlapping single-photon amplitudes would give rise to
instantaneously non-zero exchange terms that would integrate to zero over a
realistic detection time. It would be possible to observe interference
between photons in different frequency ranges provided that the detection
window were sufficiently short, but in practice no interference would be
observed regardless of the spatial configuration of the photon wavefunctions
in the region of detection. In contrast, if the photons have the same peak
frequency, the spatial overlap of single-photon amplitudes in the detection
region is sufficient to give rise to exchange terms that do not integrate to
zero over a realistic detection time. In the subsequent discussion we assume
that we are dealing with photons that have the same peak frequency, and thus
where there are overlapping single-photon amplitudes in the detection region
there are exchange effects in coincidence measurements.

We now apply this formalism to several important two-photon interference
experiments, and demonstrate that in each case the interference can be
understood as single-photon wave function amplitudes giving rise to an
exchange effect.

\subsection{The Hong-Ou-Mandel interferometer}

First we consider the simple case of the Hong-Ou-Mandel interferometer. A
schematic diagram of this experiment is shown in Fig. 1, where we label the
input ports $a$ and $b$, and the output ports $c$ and $d$. Two photons are
incident on the 50/50 beam splitter, with one of the photons delayed in time
by $\delta t$. We use a simplified notation, taking only one linear
polarization associated with each port, and considering small wave packets
with a narrow spread of wave vector components perpendicular to the
direction identified by the port. Hence a one-dimensional treatment is
possible for each port. We use, for example, $a_{b}^{\dagger }\left(
k\right) $ to denote a creation operator associated with a wave vector of
magnitude $k$ and in the propagation direction relevant for port $b$. In
this shorthand the initial separable two-photon state, at a time just before
the photon wave functions impinge on the beam-splitter, can be written as

\begin{equation}
\left\vert P_{a},Q_{b}\right\rangle =\int \int dkdk^{\prime
}g^{P}(k)g^{Q}(k^{\prime })a_{a}^{\dagger }\left( k\right) a_{b}^{\dagger
}\left( k^{\prime }\right) \left\vert vac\right\rangle ,  \label{homstate}
\end{equation}%
where we have gone to a continuous range of wave numbers. In the full
notation used in the previous section the functions $g^{P}({\bf k})$ and $%
g^{Q}({\bf k})$ would of course be orthogonal, since the single-photon wave
functions do not overlap. But in our shorthand notation we take photons $%
P_{a}$ and $Q_{b}$ to be the same when referenced to their own ports, with
only the second delayed by a time $\delta t$ from the first; thus $%
g^{Q}(k)=g^{P}(k)e^{i\omega _{k}\delta t}$. In order to reduce the
calculation to simple integration over one-dimensional coordinates we use
the scalars, $x_{n}$, and mode indices, $j_{n}$, where $x_{n}$ is the
spatial coordinate for the input port $j_{n}$, together to identify a
position ${\bf r}$ of interest. With this notation and Eq. (\ref{homstate})
the two-photon wave function in the region just before the beam splitter can be written as 
\begin{equation}
\Phi (j_{1},x_{1};j_{2},x_{2};t_{1},t_{2})=\frac{1}{\sqrt{2}}\left[ \Phi
^{P}(j_{1},x_{1};t_{1})\Phi ^{Q}(j_{2},x_{2};t_{2})+\Phi
^{Q}(j_{1},x_{1};t_{1})\Phi ^{P}(j_{2},x_{2};t_{2})\right] ,
\label{homwavefunction}
\end{equation}%
where the single-photon wavefunction $\Phi ^{P}(j,x;t)$ is defined by the
obvious simplification of (\ref{eq:singlewavefunction}) and 
\begin{eqnarray}
\Phi ^{P}(a,x;t) &=&\int dkg^{P}(k)e^{i(kx-\omega _{k}t)}=V^{P}(x,t),
\label{homwavefunction2} \\
\Phi ^{Q}(b,x;t) &=&\int dkg^{Q}(k)e^{i(kx-\omega _{k}t)}=V^{Q}(x,t), 
\nonumber \\
\Phi ^{P}(b,x;t) &=&\Phi ^{Q}(a,x;t)=0.  \nonumber
\end{eqnarray}%
The origins of ports $a$ and $b$ are equally distant from the beam splitter,
as shown in Fig. 1. We have given the one-dimensional single-photon wave
functions in this input region a special label, $V^{P(Q)}(x,t),$ so that we
can illustrate how these input wave functions interfere in the detection
region. The wave function amplitude corresponding to a two-photon detection
event in the input region before the beam splitter is

\begin{eqnarray}
\Phi (a,x_{1};b,x_{2};t_{1},t_{2}) &=&\Phi (b,x_{2};a,x_{1};t_{2},t_{1})=%
\frac{1}{\sqrt{2}}\left[ V^{P}(x_{1},t_{1})V^{Q}(x_{2},t_{2})\right]
\label{homwavefunction3} \\
&=&\frac{1}{\sqrt{2}}\left[ V^{P}(x_{1},t_{1})V^{P}(x_{2},t_{2}-\delta t)%
\right]  \nonumber
\end{eqnarray}%
and, as one would expect, the square modulus of this function does not give
rise to exchange terms and there is no interference in this region. We now
wish to evaluate the two-photon detection rate in the output ports $c$ and $%
d $; the origin of each of these ports is taken at the beam splitter. As is
well known, the effect of the beam splitter is to effect a canonical
transformation on the port operators\cite{BACHOR1},

\begin{eqnarray}
a_{a}^{\dagger }\left( k\right) &\rightarrow &\frac{e^{i\omega _{k}\Delta }}{%
\sqrt{2}}\left[ a_{d}^{\dagger }\left( k\right) -ia_{c}^{\dagger }\left(
k\right) \right]  \label{homaction} \\
a_{b}^{\dagger }\left( k\right) &\rightarrow &\frac{e^{i\omega _{k}\Delta }}{%
\sqrt{2}}\left[ a_{c}^{\dagger }\left( k\right) -ia_{d}^{\dagger }\left(
k\right) \right] ,  \nonumber
\end{eqnarray}%
where $\Delta $ is the time required for light to travel from the origins
associated with ports $(a,b)$ to the beam-splitter, and it will be
convenient below to use delayed time coordinates $\overline{t}%
_{1(2)}=t_{1(2)}-\Delta $. The two-photon wave function in the region of the
detectors has the following components%
\begin{eqnarray}
\Phi (c,x_{1};c,x_{2};t_{1},t_{2}) &=&\frac{1}{2\sqrt{2}}\left[ \Phi
^{P}(c,x_{1};\overline{t}_{1})\Phi ^{Q}(c,x_{2};\overline{t}_{2})+\Phi
^{Q}(c,x_{1};\overline{t}_{1})\Phi ^{P}(c,x_{2};\overline{t}_{2})\right] 
\nonumber \\
\Phi (d,x_{1};d,x_{2};t_{1},t_{2}) &=&\frac{1}{2\sqrt{2}}\left[ \Phi
^{P}(d,x_{1};\overline{t}_{1})\Phi ^{Q}(d,x_{2};\overline{t}_{2})+\Phi
^{Q}(d,x_{1};\overline{t}_{1})\Phi ^{P}(d,x_{2};\overline{t}_{2})\right]
\label{HOMfullwavefunction2} \\
\Phi (c,x_{1};d,x_{2};t_{1},t_{2}) &=&\Phi (d,x_{2};c,x_{1};t_{2},t_{1})=%
\frac{1}{2\sqrt{2}}\left[ \Phi ^{P}(c,x_{1};\overline{t}_{1})\Phi
^{Q}(d,x_{2};\overline{t}_{2})+\Phi ^{Q}(c,x_{1};\overline{t}_{1})\Phi
^{P}(d,x_{2};\overline{t}_{2})\right] .  \nonumber
\end{eqnarray}%
We can write the wave function in terms of the single-photon input amplitude
since%
\begin{eqnarray}
\Phi ^{P}(c,x;t) &=&-iV^{P}(x,t)  \nonumber \\
\Phi ^{P}(d,x;t) &=&V^{P}(x,t)  \label{intermsofinput} \\
\Phi ^{Q}(c,x;t) &=&V^{Q}(x,t)=V^{P}(x,t-\delta t)  \nonumber \\
\Phi ^{Q}(d,x;t) &=&-iV^{Q}(x,t)=-iV^{P}(x,t-\delta t)  \nonumber
\end{eqnarray}

and we have

\begin{eqnarray}
\Phi (c,x_{1};c,x_{2};t_{1},t_{2}) &=&\Phi (d,x_{1};d,x_{2};t_{1},t_{2})=%
\frac{-i}{2\sqrt{2}}\left[ V^{Q}(x_{1},\overline{t}_{1})V^{P}(x_{2},%
\overline{t}_{2})+V^{P}(x_{1},\overline{t}_{1})V^{Q}(x_{2},\overline{t}_{2})%
\right]  \label{HOMamplitude13} \\
\Phi (c,x_{1};d,x_{2};t_{1},t_{2}) &=&\Phi (d,x_{2};c,x_{1};t_{2},t_{1})=%
\frac{1}{2\sqrt{2}}\left[ V^{Q}(x_{1},\overline{t}_{1})V^{P}(x_{2},\overline{%
t}_{2})-V^{P}(x_{1},\overline{t}_{1})V^{Q}(x_{2},\overline{t}_{2})\right] .
\label{HOMamplitude132}
\end{eqnarray}%
One can see how the input amplitudes for the individual photons interfere
when the two-photon wave function is written in terms of the input wave
functions. As per the usual discussion, each photon is either reflected or
transmitted by the beam splitter giving four possible \textquotedblleft
outcomes\textquotedblright . For each outcome there is a corresponding wave
function component in equations (\ref{HOMamplitude13})-(\ref{HOMamplitude132}%
). The $\Phi (c,x_{1};c,x_{2};t_{1},t_{2})$ and $\Phi
(d,x_{1};d,x_{2};t_{1},t_{2})$ components are the amplitudes for a detection
event when both detectors are in the same output port. \ The amplitudes in (%
\ref{HOMamplitude132}) are those relevant to this experiment since we are
interested in the coincidence detection rate where one photon is detected in
each output port. The negative sign in (\ref{HOMamplitude132}) is the
important feature that allows for destructive interference. A $-i$ phase
shift occurs when a photon wavefunction is reflected and the amplitude
corresponding to a detection where both photons are reflected by the beam
splitter accumulates a $\pi $ phase shift relative to the amplitude of both
photons being transmitted. The coincidence detection rate of the detectors
is then proportional to

\begin{eqnarray}
\left\vert \Phi (c,x_{1};d,x_{2};t_{1},t_{2})\right\vert ^{2} &=&\frac{1}{8}%
\bigg[\left\vert V^{P}(x_{1},\overline{t}_{1})\right\vert ^{2}\left\vert
V^{P}(x_{2},\overline{t}_{2}-\delta t)\right\vert ^{2}+\left\vert
V^{P}(x_{1},\overline{t}_{1}-\delta t)\right\vert ^{2}\left\vert V^{P}(x_{2},%
\overline{t}_{2})\right\vert ^{2}  \nonumber \\
&&{}-V^{P}(x_{1},\overline{t}_{1})\left[ V^{P}(x_{1},\overline{t}_{1}-\delta
t)\right] ^{\ast }V^{P}(x_{2},\overline{t}_{2}-\delta t)\left[ V^{P}(x_{2},%
\overline{t}_{2})\right] ^{\ast }  \label{HOMrate} \\
&&{}-\left[ V^{P}(x_{1},\overline{t}_{1})\right] ^{\ast }V^{P}(x_{1},%
\overline{t}_{1}-\delta t)\left[ V^{P}(x_{2},\overline{t}_{2}-\delta t)%
\right] ^{\ast }V^{P}(x_{2},\overline{t}_{2})\bigg]{}.  \nonumber
\end{eqnarray}%
The last two terms of Eq. (\ref{HOMrate}) are the exchange terms
corresponding to the second two terms of (\ref{eq:exchangefec}). As the $P$
and $Q$ (delayed) photon amplitudes overlap in the detection region, the
exchange terms bring the detection rate to zero. At zero delay there is
maximum interference and the coincidence detection rate is zero: The
exchange terms completely cancel out the classical independent-particle
amplitude for a two-photon detection. \ Similarily, evaluating

\begin{eqnarray}
\left\vert \Phi (c,x_{1};c,x_{2};t_{1},t_{2})\right\vert ^{2} &=&\frac{1}{8}%
\bigg[\left\vert V^{P}(x_{1},\overline{t}_{1})\right\vert ^{2}\left\vert
V^{P}(x_{2},\overline{t}_{2}-\delta t)\right\vert ^{2}+\left\vert
V^{P}(x_{1},\overline{t}_{1}-\delta t)\right\vert ^{2}\left\vert V^{P}(x_{2},%
\overline{t}_{2})\right\vert ^{2}  \nonumber \\
&&{}+V^{P}(x_{1},\overline{t}_{1})\left[ V^{P}(x_{1},\overline{t}_{1}-\delta
t)\right] ^{\ast }V^{P}(x_{2},\overline{t}_{2}-\delta t)\left[ V^{P}(x_{2},%
\overline{t}_{2})\right] ^{\ast }  \label{HOMRATE2} \\
&&{}+\left[ V^{P}(x_{1},\overline{t}_{1})\right] ^{\ast }V^{P}(x_{1},%
\overline{t}_{1}-\delta t)\left[ V^{P}(x_{2},\overline{t}_{2}-\delta t)%
\right] ^{\ast }V^{P}(x_{2},\overline{t}_{2})\bigg]{}  \nonumber
\end{eqnarray}%
shows the constructive interference exchange effect that arises when
measuring the coincidence detection rate of two detectors in the same output
port. \ Assuming the photons are of finite temporal width, for large delay $\delta t$ we find $%
\left\vert \Phi (c,x_{1};c,x_{2};t_{1},t_{2})\right\vert ^{2}=\left\vert
\Phi (c,x_{1};d,x_{2};t_{1},t_{2})\right\vert ^{2}.$ This may at first sight be surprising.
According to the beam splitter transformations, each photon has a probability of $0.5$ of being
detected in port $c$, and a probability of $0.5$ of being detected in port $d$.  Since for large
$\delta t$ we expect the photons to be independent, this would lead to a probability of $0.5$ that the photons
are detected in separate exit ports, and a probability of only $0.25$ that they would both be detected in port $c$, for example.
However, it is important to recall that we have based the
theory on the Glauber detection model where it is assumed that the
interaction between the radiation field and detector is weak. When two detectors are located in the
same port this allows for each photon in that port to interact with both
detectors, in some sense double-counting the photons. \ This is the standard
result of calculations based on the Glauber detection model.

\smallskip We now expand the discussion to include the non-separable case,
where we allow for the possibility of \textquotedblleft frequency
entanglement\textquotedblright . In the frequency entangled case Eq. (\ref%
{eq:twophotwavesep}) does not apply, and one cannot separately address the
amplitudes of individual photons as previously discussed. Nonetheless, the
photon wave function formalism can be used to describe frequency entangled
two-photon interference in a slightly different way. Consider a two-port
single-polarization gaussian entangled two-photon state as an input for the
Hong-Ou-Mandel interferometer: 
\begin{eqnarray}
\left\vert P_{a},Q_{b}\right\rangle _{C} &=&N\left( \sigma \right) \int \int
dkdk^{\prime }g^{P}(k)g^{Q}(k^{\prime })e^{-\frac{\left( \omega _{k}+\omega
_{k^{\prime }}-2\omega _{k_{o}}\right) ^{2}}{\sigma ^{2}}}a_{a}^{\dagger
}\left( k\right) a_{b}^{\dagger }\left( k^{\prime }\right) \left\vert
vac\right\rangle   \label{correlatedtwophoton} \\
&=&N\left( \sigma \right) \frac{\sigma }{2\sqrt{\pi }}\int dte^{-\frac{%
\sigma ^{2}t^{2}}{4}}e^{-2i\omega _{k_{o}}t}\left\vert \Psi \left( t\right)
\right\rangle   \label{correlatedtwophotonTT}
\end{eqnarray}%
where $\sigma $ is the entanglement width,%
\begin{eqnarray}
\left\vert \Psi \left( t\right) \right\rangle  &=&\int \int dkdk^{\prime
}g^{P}(k)e^{i\omega _{k}t}g^{Q}(k^{\prime })e^{i\omega _{k^{\prime
}}t}a_{a}^{\dagger }\left( k\right) a_{b}^{\dagger }\left( k^{\prime
}\right) \left\vert vac\right\rangle ,  \label{psidisplaced} \\
N\left( \sigma \right)  &=&\left[ \int \int dkdk^{\prime }\left\vert
g^{P}(k)\right\vert ^{2}\left\vert g^{Q}(k^{\prime })\right\vert ^{2}e^{-%
\frac{2\left( \omega _{k}+\omega _{k^{\prime }}-2\omega _{k_{o}}\right) ^{2}%
}{\sigma ^{2}}}\right] ^{-\frac{1}{2}},  \label{correlatednorm}
\end{eqnarray}%
and $2\omega _{k_{o}}$ is the peak entanglement angular frequency. One can
think of this state as a superposition of separable two-photon states (\ref%
{psidisplaced}), each with the two photons temporally displaced by a time $t$%
. In the superposition (\ref{correlatedtwophotonTT}) the amplitude of each
component contain a phase factor that varies rapidly with this displacement
time. The degree of the frequency entanglement determines the weighted
distribution of these single-photon wave functions in time. It is in this
form that one can discuss frequency entangled two-photon interference in
terms of separable single-photon amplitudes.

For the Hong-Ou-Mandel experiment with frequency entangled photons, one can
describe the interference in terms of the separable input wave function
amplitudes with a superposition of two-photon wave functions distributed in
time. The two-photon detection amplitude becomes 
\begin{equation}
\Phi (c,x_{1};d,x_{2};t_{1},t_{2})=\frac{1}{4\sqrt{2}}N\left( \sigma \right) 
\frac{\sigma }{\sqrt{\pi }}\int e^{-\frac{\sigma ^{2}t^{2}}{4}}e^{-2i\omega
_{k_{o}}t}dt\left[ V^{P}(x_{1},\overline{t}_{1}-t)V^{Q}(x_{2},\overline{t}%
_{2}-t)-V^{Q}(x_{1},\overline{t}_{1}-t)V^{P}(x_{2},\overline{t}_{2}-t)\right]
.  \label{correlatedHOM0}
\end{equation}%
Comparing with (\ref{HOMamplitude132}) one can see that the phase shift of
the reflected single-photon amplitudes once again allow for destructive
interference. The detection rate is proportional to 
\begin{eqnarray}
\left\vert \Phi (c,x_{1};d,x_{2};t_{1},t_{2})\right\vert ^{2} &=&\frac{1}{32}%
\left\vert N\left( \sigma \right) \right\vert ^{2}\frac{\sigma ^{2}}{\pi }%
\int \int e^{-\frac{\sigma ^{2}\left( t+T\right) ^{2}}{4}}e^{-2i\omega
_{k_{o}}\left( t-T\right) }dtdT  \nonumber \\
&&\bigg[\left[ V^{P}(x_{1},\overline{t}_{1}-t)\right] ^{\ast }V^{P}(x_{1},%
\overline{t}_{1}-T)\left[ V^{Q}(x_{2},\overline{t}_{2}-t)\right] ^{\ast
}V^{Q}(x_{2},\overline{t}_{2}-T)  \nonumber \\
&&+\left[ V^{P}(x_{2},\overline{t}_{2}-t)\right] ^{\ast }V^{P}(x_{2},%
\overline{t}_{2}-T)\left[ V^{Q}(x_{1},\overline{t}_{1}-t)\right] ^{\ast
}V^{Q}(x_{1},\overline{t}_{1}-T)  \label{correlatedHOM} \\
&&-\left[ V^{P}(x_{1},\overline{t}_{1}-t)\right] ^{\ast }V^{Q}(x_{1},%
\overline{t}_{1}-T)\left[ V^{Q}(x_{2},\overline{t}_{2}-t)\right] ^{\ast
}V^{P}(x_{2},\overline{t}_{2}-T)  \nonumber \\
&&{}-V^{P}(x_{1},\overline{t}_{1}-T)\left[ V^{Q}(x_{1},\overline{t}_{1}-t)%
\right] ^{\ast }V^{Q}(x_{2},\overline{t}_{2}-T)\left[ V^{P}(x_{2},\overline{t%
}_{2}-t)\right] ^{\ast }\bigg]{}.  \nonumber
\end{eqnarray}%
\newline
Here we have the possibility of exchange interference between amplitudes at
different temporal displacements. Using this form it is possible to
investigate the effect of frequency-entanglement on few-photon interference
by examining the exchange terms. We will show that for a Hong-Ou-Mandel
interferometer with a large enough delay such that the exchange terms are
zero (no interference) with separable input photons, no amount of
frequency-entanglement can introduce interference. We will begin with the
assumption that the input wave functions can be well approximated as having
finite temporal width, that is, if $\left\vert a\right\vert >\beta $ then $%
V^{P}(x,t)V^{P}(x,t+a)=0,$ where $\beta $ is a photon width parameter. Given 
$\left\vert \delta t\right\vert >\beta $, the exchange terms in (\ref%
{HOMrate}) are zero and there is no interference in the separable case. To
show that in this case there is no interference between temporally displaced
two-photon wavefunctions, from examining the exchange terms in (\ref%
{correlatedHOM}) we see we must show that for all $(t,T)$%
\begin{equation}
\left[ V^{P}(x_{1},t_{1}+t)\right] ^{\ast }V^{P}(x_{1},t_{1}+\delta t+T)%
\left[ V^{P}(x_{2},t_{2}+\delta t+t)\right] ^{\ast }V^{P}(x_{2},t_{2}+T)=0.
\label{exchangeproof1}
\end{equation}%
The proof is by contradiction. First we suppose that Eq. (\ref%
{exchangeproof1}) is does not hold. \ With the finite temporal width
assumption, this implies both $\left\vert \delta t+(T-t)\right\vert <\beta $
and $\left\vert \delta t-(T-t)\right\vert <\beta .$ But this is impossible
to satisfy since $\left\vert \delta t\right\vert >\beta $. Hence (\ref%
{exchangeproof1}) must hold and regardless of the entanglement, the exchange
terms are zero if the initial relative photon delay, $\delta t$, is larger
than temporal width of the single-photon wave functions. One can see this in
Fig. 2, which shows a pictorial representation of the exchange terms in (%
\ref{correlatedHOM}). In order for the exchange terms to contribute to the
coincidence detection rate, all four single-photon wave functions must
overlap at some value of $(t,T)$. Regardless of the temporal displacements,
the single photon amplitudes in the exchange terms do not overlap for $%
\delta t$ larger than temporal width of the photon wave functions. We will
discuss later how this is not the case for the postponed compensation
experiment, where the frequency-entanglement is more significant.

\subsection{Quantum eraser experiments}

The version of the quantum eraser\cite{SCULLY1} we discuss here illustrates
how the photon wave function formalism deals with multiple polarizations.
The basic idea of the two-photon quantum eraser is as follows: It is
possible to introduce distinguishing information in the input of an
interferometer that would destroy second-order interference if the
information were not \textquotedblleft erased\textquotedblright\ at some
stage before detection\cite{QE}. In our context, erasing distinguishing
information is equivalent to producing overlapping wave functions in the
detection region.

Fig. 3 is a schematic diagram of a basic quantum eraser interferometer. Two
orthogonally polarized photons in ports (a,b) interact with a beam splitter
similar to the Hong-Ou-Mandel scenario. Since the photons are orthogonally
polarized, it is possible to distinguish the final states of the two paths
leading to a coincidence detection in ports c and d. It is the presence of
the 45 degree polarizing beam splitters in front of the detectors that
causes the distinguishing information to be destroyed so that second-order
interference can occur. The input state is described by

\begin{equation}
\left\vert P_{aH},Q_{bV}\right\rangle =\int \int dkdk^{\prime
}g^{P}(k)g^{Q}(k^{\prime })a_{aH}^{\dagger }\left( k\right) a_{bV}^{\dagger
}\left( k^{\prime }\right) \left\vert vac\right\rangle  \label{qe1}
\end{equation}%
where, for example, $a_{aH}^{\dagger }\left( k\right) $ denotes a creation
operator associated with a horizontally polarized photon of wave vector
magnitude $k$ in the propagation direction relevant for port $a$. As in the
previous subsection, we define the input region wave functions. The
two-photon wave function in the region just before the beam splitter has two
components: 
\begin{equation}
\Phi _{HV}(a,x_{1};b,x_{2};t_{1},t_{2})=\Phi
_{VH}(b,x_{2};a,x_{1};t_{2},t_{1})=\frac{1}{\sqrt{2}}%
V^{P}(x_{1},t_{1})V^{Q}(x_{2},t_{2}),  \nonumber
\end{equation}%
{\it cf.} (\ref{homwavefunction3}). Once again the relevant action of the
beam splitter maps the input ports ($a$,$b$) to the output ports ($c$,$d$)
in the following way:

\begin{eqnarray}
a_{aH}^{\dagger }\left( k\right)  &\rightarrow &\frac{e^{i\omega _{k}\Delta }%
}{\sqrt{2}}\left[ a_{dH}^{\dagger }\left( k\right) -ia_{cH}^{\dagger }\left(
k\right) \right]   \label{qe2} \\
a_{bV}^{\dagger }\left( k\right)  &\rightarrow &\frac{e^{i\omega _{k}\Delta }%
}{\sqrt{2}}\left[ a_{cV}^{\dagger }\left( k\right) -ia_{dV}^{\dagger }\left(
k\right) \right] .  \nonumber
\end{eqnarray}%
The components of the two-photon wavefunction just beyond the beam splitter
are:

\begin{eqnarray}
\Phi _{HV}(d,x_{1};d,x_{2};t_{1},t_{2}) &=&\Phi
_{VH}(d,x_{2};d,x_{1};t_{2},t_{1})=-\frac{i}{2\sqrt{2}}V^{P}(x_{1},\overline{%
t}_{1})V^{Q}(x_{2},\overline{t}_{2})  \nonumber \\
\Phi _{HV}(c,x_{1};c,x_{2};t_{1},t_{2}) &=&\Phi
_{VH}(c,x_{2};c,x_{1};t_{2},t_{1})=-\frac{i}{2\sqrt{2}}V^{P}(x_{1},\overline{%
t}_{1})V^{Q}(x_{2},\overline{t}_{2})  \label{qe47} \\
\Phi _{HV}(c,x_{1};d,x_{2};t_{1},t_{2}) &=&\Phi
_{VH}(d,x_{2};c,x_{1};t_{2},t_{1})=-\frac{1}{2\sqrt{2}}V^{P}(x_{1},\overline{%
t}_{1})V^{Q}(x_{2},\overline{t}_{2})  \nonumber \\
\Phi _{VH}(c,x_{1};d,x_{2};t_{1},t_{2}) &=&\Phi
_{HV}(d,x_{2};c,x_{1};t_{2},t_{1})=\frac{1}{2\sqrt{2}}V^{Q}(x_{1},\overline{t%
}_{1})V^{P}(x_{2},\overline{t}_{2}),  \nonumber
\end{eqnarray}%
{\it cf.} (\ref{HOMfullwavefunction2}). To demonstrate that it is indeed the
presence of the 45 degree polarizing beam splitters that change the system
as to allow interference, we use the wave functions in this region to
calculate the coincidence detection rate of the two detectors in absence of
the polarizing beam splitters. In calculating this detection rate the square
magnitude of $\Phi _{HV}(c,x_{1};d,x_{2};t_{1},t_{2})$ and $\Phi
_{VH}(c,x_{1};d,x_{2};t_{1},t_{2})$ are added separately, since 
\begin{equation}
w^{\left( 2\right) }(c,x_{1};d,x_{2};t_{1},t_{2})=\frac{1}{8}\left\vert
V^{P}(x_{1},\overline{t}_{1})V^{Q}(x_{2},\overline{t}_{2})\right\vert ^{2}+%
\frac{1}{8}\left\vert V^{Q}(x_{1},\overline{t}_{1})V^{P}(x_{2},\overline{t}%
_{2})\right\vert ^{2}.  \label{qe5}
\end{equation}%
No interference occurs if one removes the polarizing beam splitters since
there are no exchange terms in (\ref{qe5}). Now consider the experiment as
shown. The effect of the 45 degree polarizing beam splitters is to map the
ports (c,d) to the ports (e,f) as follows:

\begin{eqnarray}
a_{cV}^{\dagger }\left( k\right)  &\rightarrow &\frac{e^{i\omega _{k}\Delta
^{\prime }}}{\sqrt{2}}\left[ a_{e+}^{\dagger }\left( k\right)
+a_{g-}^{\dagger }\left( k\right) \right] ,a_{cH}^{\dagger }\left( k\right)
\rightarrow \frac{e^{i\omega _{k}\Delta ^{\prime }}}{\sqrt{2}}\left[
a_{e+}^{\dagger }\left( k\right) -a_{g-}^{\dagger }\left( k\right) \right] 
\label{qe3} \\
a_{dV}^{\dagger }\left( k\right)  &\rightarrow &\frac{e^{i\omega _{k}\Delta
^{\prime }}}{\sqrt{2}}\left[ a_{f+}^{\dagger }\left( k\right)
+a_{h-}^{\dagger }\left( k\right) \right] ,a_{dH}^{\dagger }\left( k\right)
\rightarrow \frac{e^{i\omega _{k}\Delta ^{\prime }}}{\sqrt{2}}\left[
a_{f+}^{\dagger }\left( k\right) -a_{h-}^{\dagger }\left( k\right) \right] ,
\nonumber
\end{eqnarray}%
where $\Delta $ and $\Delta ^{\prime }$ are the time intervals between the
origins of the ports, and it is convenient to use delayed time coordinates $%
\overline{t}_{1(2)}=t_{1(2)}-\Delta $, $\widetilde{t}_{1(2)}=\overline{t}%
_{1(2)}-\Delta ^{\prime }$. Here we have used the $+$ and $-$ symbols to
denote the $+45$ degree and $-45$ degree polarization bases respectively. 
The two-photon wave function component relevant to coincidence detection is%
\begin{equation}
\Phi _{++}(e,x_{1};f,x_{2};t_{1},t_{2})=\Phi
_{++}(f,x_{2};e,x_{1};t_{2},t_{1})=\frac{1}{4\sqrt{2}}\left[ V^{Q}(x_{1},%
\widetilde{t}_{1})V^{P}(x_{2},\widetilde{t}_{2})-V^{P}(x_{1},\widetilde{t}%
_{1})V^{Q}(x_{2},\widetilde{t}_{2})\right]   \label{qe6}
\end{equation}%
which is the same form as the relevant wave function component in the
detection region of the Hong-Ou-Mandel interferometer. Given that the
photons initially have the same spectral decomposition and no relative
delay, $g^{P}(k)=g^{Q}(k)$, and there is complete destructive interference
and the coincidence detection rate of the detectors shown is zero as $\Phi
_{++}(e,x_{1};f,x_{2};t_{1},t_{2})=0$.

\subsection{The postponed compensation experiment}

The postponed compensation experiment was originally performed by Pittman 
{\it et al.}\cite{PITTMAN1}. The authors claim to have demonstrated an
interference effect with two photons which do not arrive simultaneously at
the beam splitter in an unbalanced Hong-Ou-Mandel type interferometer. They
conclude that this effect can not be described in terms of the overlap of
the individual photon wave packets on a beam splitter, and hence
\textquotedblleft two-photon interference\textquotedblright\ can not be
considered as the \textquotedblleft interference of two
photons\textquotedblright . As we show below, the approach we have
introduced here allows for a more precise formulation of such statements,
thus both clearly identifying the physical insight they express and
evaluating their validity.

Fig. 4 shows the schematic diagram for the postponed compensation
experiment. The input state consists of orthogonally polarized photons, with
the vertically polarized photon in port b delayed by time $\tau _{1}$ with
respect to the horizontally polarized photon in port a. A relative delay, $%
\tau _{2}$, is introduced in the horizontally polarized mode in the right
arm of the interferometer. When $\tau _{2}=2\tau _{1}$ the second delay
compensates for the initial vertical photon delay in such a way as to create
maximum interference for frequency entangled photons. While frequency
entangled photons are used as the input state in this experiment, in order
to understand this interferometer in the photon wave function picture we
first suppose that one begins with the separable input state

\begin{equation}
\left\vert P_{aH},P_{bV}\right\rangle =\int \int dkdk^{\prime
}g^{P}(k)g^{P}(k^{\prime })a_{aH}^{\dagger }\left( k\right) a_{bV}^{\dagger
}\left( k^{\prime }\right) \left\vert vac\right\rangle  \label{pc1}
\end{equation}%
for which the two-photon input wave function is given by

\begin{equation}
\Phi _{HV}(a,x_{1};b,x_{2};t_{1},t_{2})=\Phi
_{VH}(b,x_{2};a,x_{1};t_{2},t_{1})=\frac{1}{\sqrt{2}}\left[
V^{P}(x_{1},t_{1})V^{P}(x_{2},t_{2})\right] .  \label{pc1p5}
\end{equation}%
The relevant action of the optics is to map the input ports in the following
way

\begin{eqnarray}
a_{aH}^{\dagger }\left( k\right) &\rightarrow &\frac{e^{i\omega _{k}\Delta }%
}{2}\left[ \left\{ a_{d+}^{\dagger }\left( k\right) -a_{f-}^{\dagger }\left(
k\right) \right\} e^{i\omega _{k}\tau _{2}}-i\left\{ a_{c+}^{\dagger }\left(
k\right) -a_{e-}^{\dagger }\left( k\right) \right\} \right]  \label{pc2} \\
a_{bV}^{\dagger }\left( k\right) &\rightarrow &\frac{e^{i\omega _{k}\Delta }%
}{2}\left[ \left\{ a_{c+}^{\dagger }\left( k\right) +a_{e-}^{\dagger }\left(
k\right) \right\} e^{i\omega _{k}\tau _{1}}-i\left\{ a_{d+}^{\dagger }\left(
k\right) +a_{f-}^{\dagger }\left( k\right) \right\} e^{i\omega _{k}\tau _{1}}%
\right] ,  \nonumber
\end{eqnarray}%
which leads to the relevant two-photon wave function component in the region
of the detectors

\begin{equation}
\Phi _{++}(c,x_{1};d,x_{2};t_{1},t_{2})=\frac{1}{4\sqrt{2}}\left[
V^{P}(x_{2},\overline{t}_{2}-\tau _{2})V^{P}(x_{1},\overline{t}_{1}-\tau
_{1})-V^{P}(x_{1},\overline{t}_{1})V^{P}(x_{2},\overline{t}_{2}-\tau _{1})%
\right] ,  \label{pc3}
\end{equation}%
where $\overline{t}_{1(2)}=t_{1(2)}-\Delta .$ When the interferometer is
adjusted so that, $\tau _{2}=2\tau _{1}$, one can see that the separable
photon input yields no interference. The coincidence detection rate at
maximum interference is proportional to

\begin{eqnarray}
\left\vert \Phi _{++}(c,x_{1};d,x_{2};t_{1},t_{2})\right\vert ^{2} &=&\frac{1%
}{32}\bigg[\left\vert V^{P}(x_{2},\overline{t}_{2}-2\tau _{1})\right\vert
^{2}\left\vert V^{P}(x_{1},\overline{t}_{1}-\tau _{1})\right\vert
^{2}+\left\vert V^{P}(x_{1},\overline{t}_{1})\right\vert ^{2}\left\vert
V^{P}(x_{2},\overline{t}_{2}-\tau _{1})\right\vert ^{2}  \nonumber \\
&&-\left[ V^{P}(x_{2},\overline{t}_{2}-2\tau _{1})V^{P}(x_{1},\overline{t}%
_{1}-\tau _{1})\right] ^{\ast }V^{P}(x_{1},\overline{t}_{1})V^{P}(x_{2},%
\overline{t}_{2}-\tau _{1})  \label{pc4} \\
&&-V^{P}(x_{2},\overline{t}_{2}-2\tau _{1})V^{P}(x_{1},\overline{t}_{1}-\tau
_{1})\left[ V^{P}(x_{1},\overline{t}_{1})V^{P}(x_{2},\overline{t}_{2}-\tau
_{1})\right] ^{\ast }\bigg]{}  \nonumber
\end{eqnarray}%
and the exchange terms are always zero, assuming the time delay $\tau _{1}$
is much greater than the width of the wave functions. With the separable
input state there are indeed no overlapping single photon amplitudes at the
beam splitter, as Pittman {\it et al.} claim. However, one also does not
observe any interference!

It is only when the input photons are frequency entangled that one can
measure an interference effect. Consider the frequency entangled input state 
\begin{equation}
\left\vert P_{aH},P_{bV}\right\rangle _{C}=N\left( \sigma \right) \int \int
dkdk^{\prime }g^{P}(k)g^{P}(k^{\prime })e^{-\frac{\left( \omega _{k}+\omega
_{k^{\prime }}-2\omega _{k_{o}}\right) ^{2}}{\sigma ^{2}}}a_{aH}^{\dagger
}\left( k\right) a_{bV}^{\dagger }\left( k^{\prime }\right) \left\vert
vac\right\rangle  \label{pc5}
\end{equation}%
which leads to the relevant two-photon wave function component

\begin{equation}
\Phi _{++}(c,x_{1};d,x_{2};t_{1},t_{2})=\frac{1}{4\sqrt{2}}N\left( \sigma
\right) \frac{\sigma }{\sqrt{\pi }}\int e^{-\frac{\sigma ^{2}t^{2}}{4}%
}e^{-2i\omega _{k_{o}}t}dt\left[ 
\begin{array}{c}
V^{P}(x_{2},\overline{t}_{2}-\tau _{2}-t)V^{P}(x_{1},\overline{t}_{1}-\tau
_{1}-t) \\ 
-V^{P}(x_{1},\overline{t}_{1}-t)V^{P}(x_{2},\overline{t}_{2}-\tau _{1}-t)%
\end{array}%
\right] .  \label{pc6}
\end{equation}%
The coincidence detection rate at maximum interference is then proportional
to

\begin{eqnarray}
\left\vert \Phi _{++}(c,x_{1};d,x_{2};t_{1},t_{2})\right\vert ^{2} &=&\frac{1%
}{32}\left\vert N\left( \sigma \right) \right\vert ^{2}\frac{\sigma ^{2}}{%
\pi }\int \int e^{-\frac{\sigma ^{2}\left( t+T\right) ^{2}}{4}}e^{-2i\omega
_{k_{o}}\left( t-T\right) }dtdT  \nonumber \\
&&\bigg[\left[ V^{P}(x_{1},\overline{t}_{1}-t)\right] ^{\ast }V^{P}(x_{1},%
\overline{t}_{1}-T)\left[ V^{P}(x_{2},\overline{t}_{2}-\tau _{1}-t)\right]
^{\ast }V^{P}(x_{2},\overline{t}_{2}-\tau _{1}-T)  \nonumber \\
&&+\left[ V^{P}(x_{2},\overline{t}_{2}-2\tau _{1}-t)\right] ^{\ast
}V^{P}(x_{2},\overline{t}_{2}-2\tau _{1}-T)\left[ V^{P}(x_{1},\overline{t}%
_{1}-\tau _{1}-t)\right] ^{\ast }V^{P}(x_{1},\overline{t}_{1}-\tau _{1}-T)
\label{pc7} \\
&&-\left[ V^{P}(x_{1},\overline{t}_{1}-t)\right] ^{\ast }V^{P}(x_{1},%
\overline{t}_{1}-\tau _{1}-T)\left[ V^{P}(x_{2},\overline{t}_{2}-\tau _{1}-t)%
\right] ^{\ast }V^{P}(x_{2},\overline{t}_{2}-2\tau _{1}-T)  \nonumber \\
&&{}-V^{P}(x_{1},\overline{t}_{1}-T)\left[ V^{P}(x_{1},\overline{t}_{1}-\tau
_{1}-t)\right] ^{\ast }V^{P}(x_{2},\overline{t}_{2}-\tau _{1}-T)\left[
V^{P}(x_{2},\overline{t}_{2}-2\tau _{1}-t)\right] ^{\ast }\bigg]{}. 
\nonumber
\end{eqnarray}%
This expression is similar to that of the Hong-Ou-Mandel interferometer with
frequency entangled photons (\ref{correlatedHOM}), but in this case the
frequency-entanglement is necessary for interference. Given that the
entanglement is sufficiently strong $(1/\sigma >>\tau _{1})$, the exchange
terms are non-zero, and one observes interference between the temporally
distributed amplitudes of the two-photon wave function. A strong enough
frequency entanglement introduces interference when there is no interference
in the equivalent separable-photon input experiment. In order to show this
we will again assume that the wave functions have a finite width: if $%
\left\vert a\right\vert >\beta $ then $V^{P}(x,t)V^{P}(x,t+a)=0,$ where $%
\beta $ is the photon width parameter. Given $\left\vert \tau
_{1}\right\vert >\beta $, the exchange terms in (\ref{pc4}) are zero and
there is no interference in the separable case. Examining the exchange terms
in (\ref{pc7}) we can observe interference in the frequency entangled case if%
\begin{equation}
\left[ V^{P}(x_{1},t_{1}+t)\right] ^{\ast }V^{P}(x_{1},t_{1}+\tau _{1}+T)%
\left[ V^{P}(x_{2},t_{2}+\tau _{1}+t)\right] ^{\ast }V^{P}(x_{2},t_{2}+2\tau
_{1}+T)\neq 0.  \label{pcproof}
\end{equation}%
With the finite width assumption Eq. (\ref{pcproof}) implies $\left\vert
\tau _{1}+(T-t)\right\vert <\beta $. Fig. 5 is a pictorial representation
of how the temporally displaced single-photon wave functions overlap in the
region of the detectors to produce non-zero exchange terms given $\left\vert
\tau _{1}+(T-t)\right\vert <\beta $.

In their discussion of this system, Pittman {\it et al.} state that the
photons arrive at the beam splitter at much different times. Perhaps there
exists some precise definition of what it means for \textquotedblleft a
photon to arrive at the beam splitter\textquotedblright\ for which this
would be true. However, if one describes the frequency entangled photons
according to the wave function picture discussed here, one finds that the
wave function amplitudes of both the signal and idler are simultanously
non-zero at the beam splitter.

From a broader perspective, of course, we can agree with Pittman {\it et al.}
that \textquotedblleft the intuitively comforting notion of the photons
overlapping at the beam splitter is not at the heart of the
interference.\textquotedblright\ One could - although Pittman {\it et al.}
did not - perform an interference experiment where single-photon amplitudes
do not overlap at a particular beam splitter (see Fig. 6). This would
demonstrate, as Pittman {\it et al.} intended, that the notion of
interference arising only when two single photons \textquotedblleft
meet\textquotedblright\ at a beam splitter is oversimplistic. Yet such an
experiment would not demonstrate any limitation of a description of
two-photon interference based on single-photon amplitudes. Even from this
broader perspective we feel Pittman {\it et al.} go too far when they claim
that \textquotedblleft two-photon interference cannot be pictured as the
interference between two single photons,\textquotedblright\ at least insofar
as it implies that there can be no general model of two-photon interference
involving single-photon amplitudes. Indeed, we have presented such a model
here. The coincidence detection rate does not explicitly depend on the
amplitudes at a particular beam splitter or any other intermediate region,
but rather the amplitudes in the detection region. We have shown in Eq. (\ref%
{eq:exchangefec}) that it is overlapping single-photon amplitudes {\it in
the detection region} that gives rise to interference.

Strekalov {\it et al.}\cite{STREKALOV1} claim there are limitations of a
single-photon wave packet approach for describing frequency entangled
two-photon interference. Using the above theory to describe their
experiment, one easily finds that the two-photon interference can in fact be
understood in terms of temporally displaced pairs of single-photon
amplitudes. In another paper, Kim {\it et al.}\cite{KIM2} observe quantum
interference between two temporally distinguishable pulses and discuss the
limitation of a single-photon amplitude description. If one examines their
experiment in terms of single-photon wave functions, one can see that the
physics is the same as the postponed compensation experiment. The difference
between the two experiments is that the temporal distribution is created by
a series of pump pulses instead of a CW pump. The 50\% visibility observed
by Kim {\it et al.} is not surprising, at least in the photon wave function
picture, as one can see it is simply postponed compensation interference
with 50\% probability. More recently, Kim and Grice\cite{KIM3} report they
have observed quantum interference in an experiment where the detected
photons retain distinguishing information. Using the photon wavefunction
theory to describe their experiment, one finds that the interfering
single-photon amplitudes are overlapping in the region of the detectors.

\section{Conclusion}

Using a particular definition of the photon wave function, we have provided
the formalism necessary for understanding second-order two-photon
intereference in terms of individual photon amplitudes. The theory clarifies
the idea of overlapping photons and replaces less rigorous explanations
involving distinguishing information or Feynman paths. We have shown that
the theory can be applied to both the separable and frequency entangled
cases, and that in the latter it allows us to eliminate some of the
confusion surrounding the interpretation of two-photon interference. This
formalism shows how photon interference can be understood as an exchange
effect by drawing an analogy to massive particle wavefunctions. \ For the
systems discussed here, the presence of exchange terms in the coincidence
detection rate expression is a necessary and sufficient condition for second
order interference. Furthermore, if one considers only detectors with
reasonable detection times and photon pairs with the same center frequency,
second order interference is equivalent to the overlap of single photon wave
functions in the detection region.

\section{Acknowledgements}

 We would like to thank K.J. Resch, J.S. Lundeen, A.M. Steinberg, and M.G. Raymer for useful discussions. This work
was supported by The Natural Sciences and Engineering Research Council of Canada.

\pagebreak

\begin{figure}
\scalebox{0.3}{\includegraphics{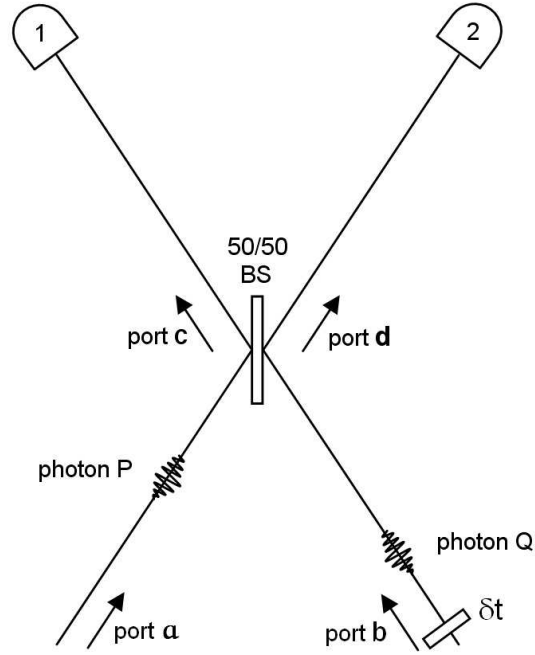}}
\caption{\label{fig:fig1} A schematic diagram of the Hong-Ou-Mandel interferometer.  Two like-polarized photons are incident on a beam splitter. The arrival of the photon in port $b$ is delayed by $\delta t$. Interference is observed in the coincidence detection rate of the two detectors. The origins of ports $c$ and $d$ are at the beam splitter, while the origins of ports $a$ and $b$ are at distance $c \Delta$ behind the beam splitter.}
\end{figure}

\begin{figure}
\scalebox{0.3}{\includegraphics{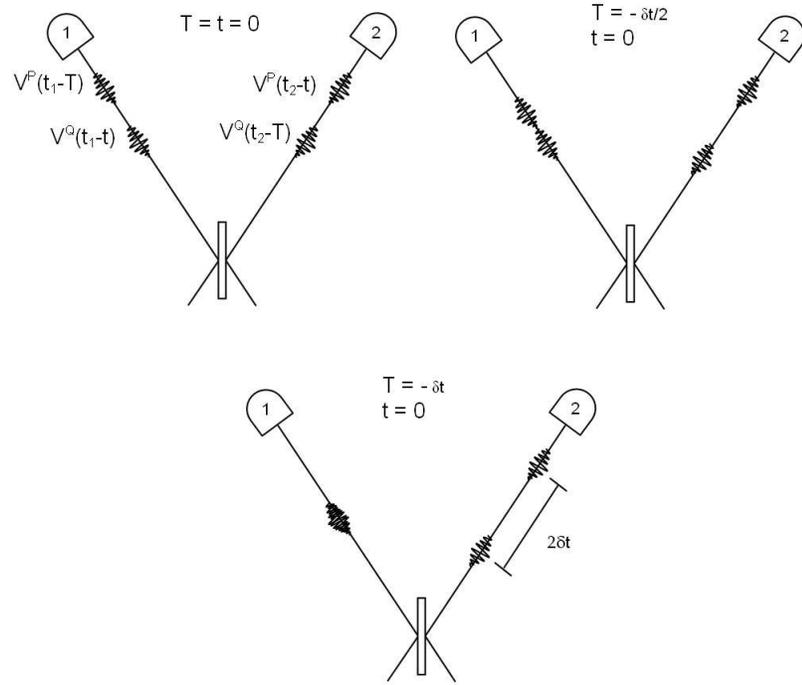}}
\caption{\label{fig:fig2} A pictorial representation of the Hong-Ou-Mandel exchange term photon wave functions in the region of the detectors. The dotted line indicates a two-photon wave function pair. In order for interference to occur, all four single photon amplitudes must overlap. Photon frequency entanglement allows for interference between temporally displaced two-photon wave functions. However, given that $\delta t$ is larger than the temporal width of the single-photon wave functions there is no interference between temporally displaced wave functions.}
\end{figure}

\begin{figure}
\scalebox{0.3}{\includegraphics{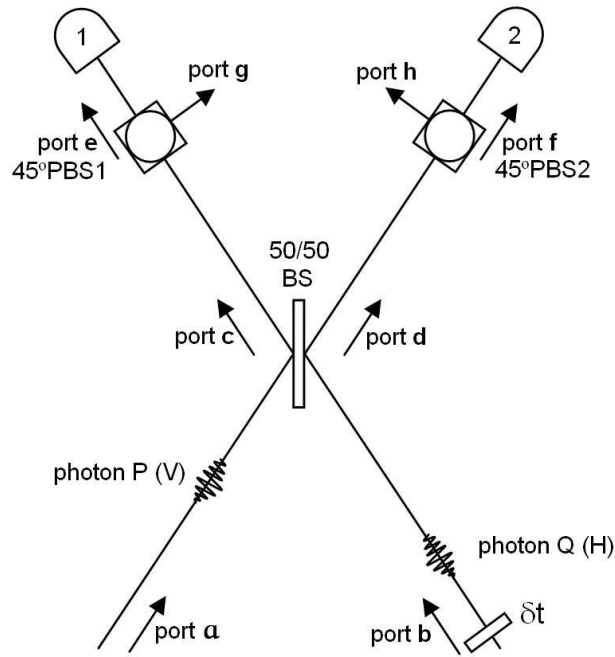}}
\caption{\label{fig:fig3} A schematic diagram of a quantum eraser.  Orthogonally polarized photons are incident on a beam splitter.  Interference in the coincidence detection rate is observed only in the region beyond the polarizing beam splitters PBS1 and PBS2.}
\end{figure}

\begin{figure}
\scalebox{0.3}{\includegraphics{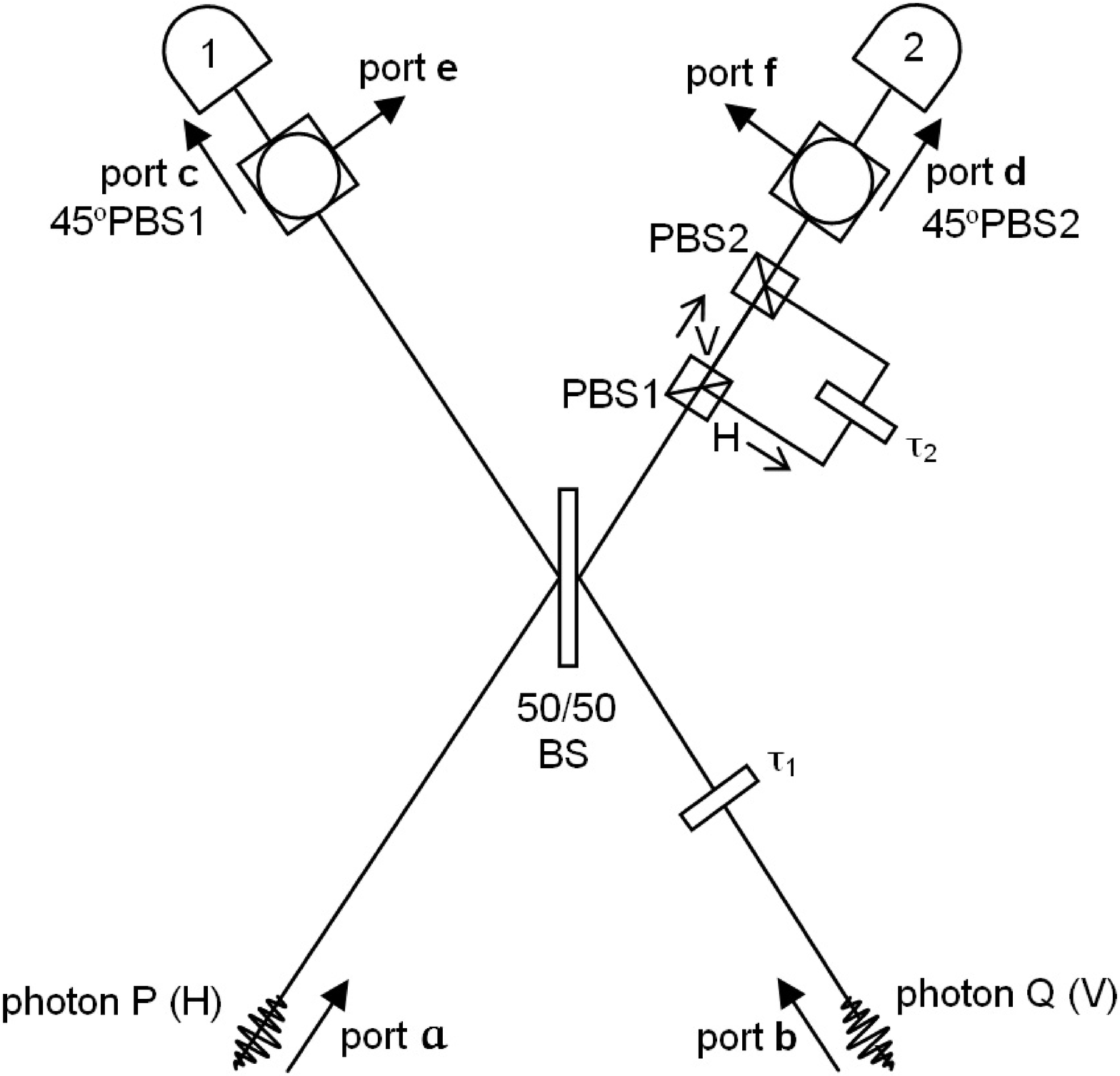}}
\caption{\label{fig:fig4} A schematic diagram of the postponed compensation experiment.  When $\tau_2 = 2 \tau_1$ interference occurs in the coincidence detection rate despite a delay in mode $b$. Frequency entangled photons are used in this experiment as no interference is observed with a separable photon state as the input.}
\end{figure}

\begin{figure}
\scalebox{0.2}{\includegraphics{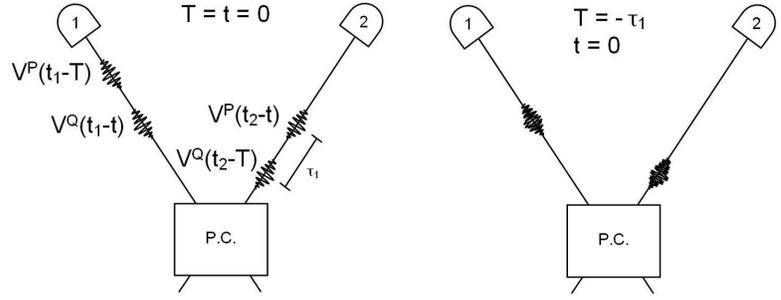}}
\caption{\label{fig:fig5} A pictorial representation postponed compensation exchange term photon wave functions in the region of the detectors. The dotted line indicates a two-photon wave function pair and P.C. denotes the postponed compensation apparatus. Once again, in order for interference to occur, all four single-photon amplitudes must overlap. No interference occurs in the separable case, but given a strong enough photon frequency entanglement the exchange terms contribute to the coincidence detection rate. One can observe interference in the frequency entangled case since the spatially displaced exchange term wave functions overlap if $\left\vert \tau_1 + (T-t)\right\vert  <\beta $, where $\beta$ is the photon width.}
\end{figure}

\begin{figure}
\scalebox{0.3}{\includegraphics{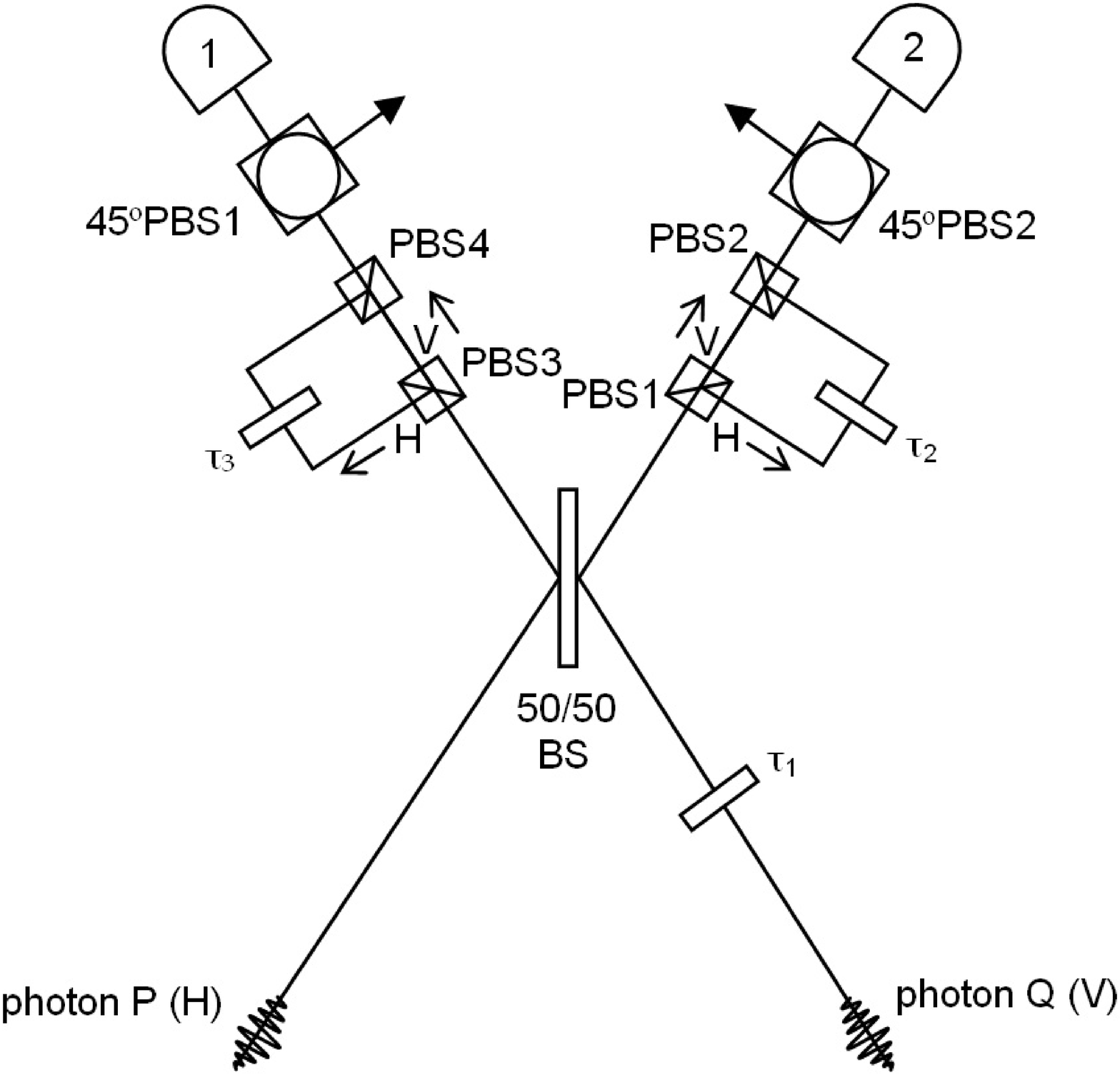}}
\caption{\label{fig:fig6} A schematic diagram of an experiment designed to observe second order interference while the photons do not meet at the central beamsplitter.  A separable photon input state is used.  With $\tau_1 = \tau_2 = \tau_3$ maximum interference occurs in the coincidence detection rate. }
\end{figure}

\end{document}